\documentclass[12pt]{article}
\usepackage{a4wide,amsmath,amssymb,graphicx,xcolor}
\numberwithin{equation}{section}
\newcommand{\1}[1]{\mathbf{1}^{#1}}
\newcommand{\2}[1]{\mathbf{2}^{#1}}
\newcommand{\3}[1]{\mathbf{3}^{#1}}

\newcommand{\5}[1]{\mathbf{5}^{#1}}
\newcommand{\F}[4]{\,_{#1}F_{#2}\left(\left.\begin{array}{c}#3\end{array}\right|#4\right)}

\begin{document}
\begin{flushright}
\large TTP12-019
\end{flushright}
\vspace{5mm}
\renewcommand{\thefootnote}{\fnsymbol{footnote}}
\begin{center}
\Large Massless two-loop self-energy diagram:\\
\large Historical review%
\footnote{Extended version of the talk at the international conference
\textit{Advances of quantum field theory},
Dubna, October 4--7, 2011.}\\[5mm]
\large A.\,G.~Grozin
\end{center}
\renewcommand{\thefootnote}{\arabic{footnote}}
\setcounter{footnote}{0}
\abstract{This class of diagrams has numerous applications.
Many interesting results have been obtained for it.}

\section{Introduction}
\label{S:Intro}

\begin{figure}[ht]
\begin{center}
\begin{picture}(124,38)
\put(28.5,19){\makebox(0,0){\includegraphics{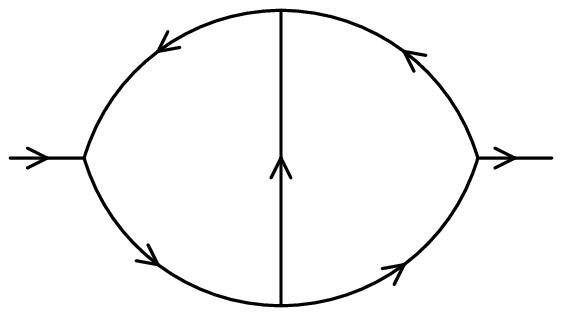}}}
\put(4.75,16){\makebox(0,0){$k$}}
\put(52.25,16){\makebox(0,0){$k$}}
\put(27.5,19){\makebox(0,0)[r]{$k_1-k_2$}}
\put(16,5){\makebox(0,0){$k_1$}}
\put(41,5){\makebox(0,0){$k_2$}}
\put(12,33){\makebox(0,0){$k_1-k$}}
\put(45,33){\makebox(0,0){$k_2-k$}}
\put(29.5,19){\makebox(0,0)[l]{3}}
\put(18,10){\makebox(0,0){5}}
\put(39,10){\makebox(0,0){4}}
\put(18,28){\makebox(0,0){1}}
\put(39,28){\makebox(0,0){2}}
\put(95.5,19){\makebox(0,0){\includegraphics{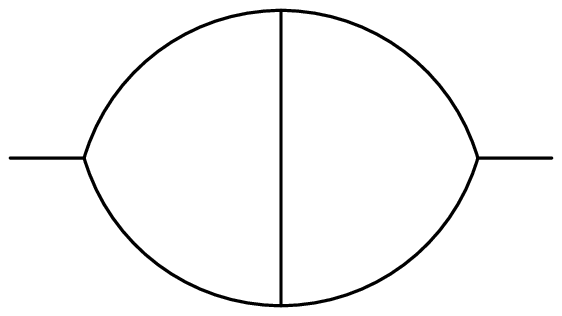}}}
\put(74.5,17){\makebox(0,0)[r]{$0$}}
\put(116.5,17){\makebox(0,0)[l]{$x$}}
\put(95.5,2){\makebox(0,0){$x_2$}}
\put(95.5,36){\makebox(0,0){$x_1$}}
\put(94.5,19){\makebox(0,0)[r]{$\bar{3}$}}
\put(83,5){\makebox(0,0){$\bar{5}$}}
\put(108,5){\makebox(0,0){$\bar{4}$}}
\put(83,33){\makebox(0,0){$\bar{1}$}}
\put(108,33){\makebox(0,0){$\bar{2}$}}
\end{picture}
\end{center}
\caption{Two-loop self-energy diagram in momentum and coordinate space.}
\label{F:Intro}
\end{figure}

We consider the integral (Fig.~\ref{F:Intro})
\begin{equation}
I(a_1,a_2,a_2,a_4,a_5) = \frac{(k^2)^{\sum a_i-d}}{\pi^d}
\int \frac{d^d k_1\,d^d k_2}%
{[(k_1-k)^2]^{a_1} [(k_2-k)^2]^{a_2} [(k_1-k_2)^2]^{a_3} (k_2^2)^{a_4} (k_1^2)^{a_5}}
\label{Intro:mom}
\end{equation}
in $d=4-2\varepsilon$-dimensional Euclidean momentum space.
It has a long and interesting history.
For many years, most of the information we had about
perturbative quantum field theory was coming (directly or indirectly)
from this integral.
All massless three-loop self-energy integrals (with integer indices)
reduce to 6 master integrals~\cite{CT:81},
5 of which are particular cases of $I$~(\ref{Intro:mom}).
Only one master integral (the non-planar one)
does not reduce to $I$;
however, using the gluing method~\cite{CT:81},
one can easily show that its value at $\varepsilon=0$
is equal to the ladder integral,
which reduces to $I(1,1,\varepsilon,1,1)$.
At four loops~\cite{BC:10}, 15 master integrals (of 28) reduce to $I$,
and thus can be easily expanded in $\varepsilon$ up to high powers.

This integral in coordinate space (Fig.~\ref{F:Intro})
\begin{align}
I(a_1,a_2,a_2,a_4,a_5) &\sim
\int \frac{d^d x_1\,d^d x_2}%
{(x_1^2)^{\bar{a}_1} [(x_1-x)^2]^{\bar{a}_2} [(x_1-x_2)^2]^{\bar{a}_3}
[(x_2-x)^2]^{\bar{a}_4} (x_2^2)^{\bar{a}_5}}
\nonumber\\
&{} \sim I(\bar{a}_2,\bar{a}_4,\bar{a}_3,\bar{a}_5,\bar{a}_1)\,,
\label{Intro:coord}
\end{align}
has the same form~(\ref{Intro:mom}) if we rename $x_i\to p_i$;
here trivial $\Gamma$-functions from Fourier transforms
are not explicitly shown, and
\begin{equation}
\bar{a}_i = \frac{d}{2} - a_i\,.
\label{Intro:dual}
\end{equation}

We can perform inversion of the integration momenta in~(\ref{Intro:mom})
\begin{equation*}
k_i = \frac{k_i'}{k_i^{\prime2}}\,,\quad
k_i^2 = \frac{1}{k_i^{\prime2}}\,,\quad
d^d k_i = \frac{d^d k_i'}{(k_i^{\prime2})^d}\,,\quad
(k_1-k_2)^2 = \frac{(k_1'-k_2')^2}{k_1^{\prime2}\,k_2^{\prime2}}\,,
\end{equation*}
and obtain
\begin{equation}
\raisebox{-9.25mm}{\begin{picture}(30,21)
\put(15,10.5){\makebox(0,0){\includegraphics{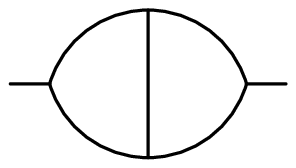}}}
\put(17,10){\makebox(0,0){$a_3$}}
\put(7,4){\makebox(0,0){$a_5$}}
\put(23,4){\makebox(0,0){$a_4$}}
\put(7,16.5){\makebox(0,0){$a_1$}}
\put(23,16.5){\makebox(0,0){$a_2$}}
\end{picture}} =
\raisebox{-9.25mm}{\begin{picture}(50,21)
\put(25,10.5){\makebox(0,0){\includegraphics{small.eps}}}
\put(27,10){\makebox(0,0){$a_3$}}
\put(10,8){\makebox(0,0){$d-a_5$}}
\put(10,4){\makebox(0,0){${}-a_1-a_3$}}
\put(40,8){\makebox(0,0){$d-a_4$}}
\put(40,4){\makebox(0,0){${}-a_2-a_3$}}
\put(17,17){\makebox(0,0){$a_1$}}
\put(33,17){\makebox(0,0){$a_2$}}
\end{picture}}\quad.
\label{Intro:inv}
\end{equation}
Inversion relations can be also derived in coordinate space, of course.

The integrals
\begin{equation*}
\raisebox{-7.25mm}{\begin{picture}(30,17)
\put(15,8.5){\makebox(0,0){\includegraphics{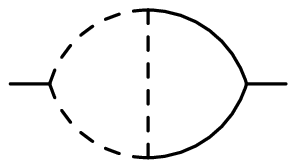}}}
\end{picture}}\,,
\end{equation*}
where dashed lines have integer indices,
have been calculated in~\cite{CKT:80}
via Gegenbauer polynomials.
Of course, now we know that it is trivial to calculate them
using IBP (Sect.~\ref{S:IBP}).

\section{Integration by parts}
\label{S:IBP}

The IBP relations for this particular class of integrals
first appeared in~\cite{VPK:81}.
They are described in the text below the formula~(15);
this formula is the homogeneity relation
(which is a consequence of the IBP relations).
Soon IBP relations evolved into a fantastically universal
and efficient method for reducing all scalar integrals of a given topology
to a few master integrals~\cite{CT:81}.

The IBP relations allow one to trivially reduce integrals
with integer indices in the left (or right) triangle
to one-loop integrals expressible via $\Gamma$-functions:
\begin{equation}
\raisebox{-7.25mm}{\begin{picture}(30,17)
\put(15,8.5){\makebox(0,0){\includegraphics{ibpa.eps}}}
\end{picture}} \to
\raisebox{-7.25mm}{\begin{picture}(30,17)
\put(15,8.5){\makebox(0,0){\includegraphics{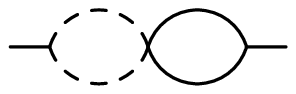}}}
\end{picture}}\,,
\raisebox{-7.25mm}{\begin{picture}(30,17)
\put(15,8.5){\makebox(0,0){\includegraphics{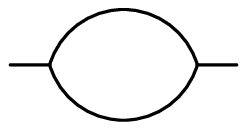}}}
\end{picture}}\,.
\label{IBP:red}
\end{equation}

If $a_3$ is not integer, things are more difficult.
The combination~\cite{CT:81} of the IBP relations
\begin{equation}
\bigl[ (d - 2 a_3 - 4) \3+ + 2 (d - a_3 -3) \bigr]
\raisebox{-4mm}{\begin{picture}(22,11)
\put(11,5.5){\makebox(0,0){\includegraphics{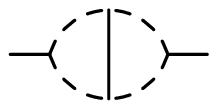}}}
\put(11.5,5){\makebox(0,0)[l]{$a_3$}}
\end{picture}}
= 2 \1+ (\5- - \2- \3+)
\raisebox{-4mm}{\begin{picture}(22,11)
\put(11,5.5){\makebox(0,0){\includegraphics{ibpd.eps}}}
\put(11.5,5){\makebox(0,0)[l]{$a_3$}}
\end{picture}}
\label{IBP:a3}
\end{equation}
allows one to shift $a_3$ by $\pm1$
(if all integer indices are 1,
all integrals in the right-hand side of~(\ref{IBP:a3})
are trivial).

\section{Uniqueness}
\label{S:Uni}

Many interesting results for massless self-energy integrals were obtained
using the method of uniqueness~\cite{VPK:81,U:83,K:84,K:85}
(see also the textbook~\cite{V:98}).
It is based on the following relations.
In coordinate space
\begin{equation}
\raisebox{-5.75mm}{\begin{picture}(22,14)
\put(11,7){\makebox(0,0){\includegraphics{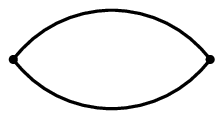}}}
\put(11,12.5){\makebox(0,0)[b]{$a_1$}}
\put(11,1.5){\makebox(0,0)[t]{$a_2$}}
\end{picture}} =
\raisebox{-5.75mm}{\begin{picture}(22,14)
\put(11,7){\makebox(0,0){\includegraphics{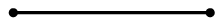}}}
\put(11,6.5){\makebox(0,0)[t]{$a_1+a_2$}}
\end{picture}}\,,\qquad
\raisebox{-2.75mm}{\begin{picture}(22,8)
\put(11,4){\makebox(0,0){\includegraphics{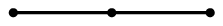}}}
\put(6,3.5){\makebox(0,0)[t]{$a_1$}}
\put(16,3.5){\makebox(0,0)[t]{$a_2$}}
\end{picture}} \sim
\raisebox{-2.75mm}{\begin{picture}(22,8)
\put(11,4){\makebox(0,0){\includegraphics{uni1.eps}}}
\put(11,3.5){\makebox(0,0)[t]{$a_1+a_1-\frac{d}{2}$}}
\end{picture}}
\label{Uni:comb}
\end{equation}
(in momentum space the second formula becomes trivial,
and the first one contains some factor;
these combinations of $\Gamma$-functions from Fourier transforms
are not explicitly shown here).
The main element of the method is the star--triangle relation
which is valid if $a_1 + a_2 + a_3 = d$:
\begin{equation}
\raisebox{-7.25mm}{\begin{picture}(19,20)
\put(9.5,11.5){\makebox(0,0){\includegraphics{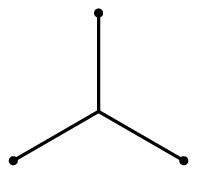}}}
\put(10,14){\makebox(0,0)[l]{$a_1$}}
\put(5.5,7){\makebox(0,0)[br]{$a_2$}}
\put(13.5,7){\makebox(0,0)[bl]{$a_3$}}
\end{picture}} \sim
\raisebox{-7.25mm}{\begin{picture}(19,20)
\put(9.5,11.5){\makebox(0,0){\includegraphics{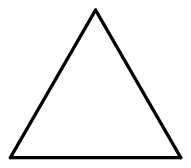}}}
\put(9.5,3){\makebox(0,0)[t]{$\bar{a}_1$}}
\put(4.6,11){\makebox(0,0)[br]{$\bar{a}_3$}}
\put(14.4,11){\makebox(0,0)[bl]{$\bar{a}_2$}}
\end{picture}}\,,
\label{Uni:st}
\end{equation}
where $\bar{a}_i$ are defined by~(\ref{Intro:dual})
(note that $\bar{a}_1 + \bar{a}_2 + \bar{a}_3 = d/2$).
It can be easily derived using inversion.

Kazakov~\cite{K:85} has calculated a non-trivial integral
\begin{equation}
I(a) = \raisebox{-4mm}{\begin{picture}(22,11)
\put(11,5.5){\makebox(0,0){\includegraphics{ibpd.eps}}}
\put(11.5,5){\makebox(0,0)[l]{$a$}}
\end{picture}}
\label{Uni:Ia}
\end{equation}
(the dashed lines have indices 1)
via hypergeometric functions of the argument $-1$%
\footnote{Earlier a particular case of this family $I(\varepsilon)$
has been calculated via hypergeometric functions of 1~\cite{H:82},
see Appendix~\ref{S:Hathrell}.}.
It has a symmetry property
\begin{equation}
I(1+a) = I(1-a-3\varepsilon)
\label{Uni:sym}
\end{equation}
(see Sect.~\ref{S:Sym});
$I(0)$ is known.
The IBP relation~(\ref{IBP:a3}) gives
\begin{equation}
I(1+a) = \frac{1-a-2\varepsilon}{a+\varepsilon} I(a)
- 2
\frac{(1-2a-3\varepsilon) \Gamma^2(1-\varepsilon) \Gamma(-a-\varepsilon) \Gamma(a+2\varepsilon)}%
{(a+\varepsilon) \Gamma(1+a) \Gamma(2-a-3\varepsilon)}\,.
\label{Uni:IBP}
\end{equation}
If we write $I(1+a)$ via a new function $G(1+a)$
\begin{equation*}
I(1+a) = 2
\frac{\Gamma^2(1-\varepsilon) \Gamma(-a-\varepsilon) \Gamma(a+2\varepsilon)}%
{\Gamma(1+a) \Gamma(1-a-3\varepsilon)}
G(1+a)\,,
\end{equation*}
then this recurrence relation becomes simpler:
\begin{equation*}
G(1+a) = \frac{a}{1-a-3\varepsilon} G(a)
+ \frac{1}{a-1+3\varepsilon}
\left( \frac{1}{a+\varepsilon} + \frac{1}{a-1+2\varepsilon} \right)\,.
\end{equation*}
Writing this function as a sum over its poles
\begin{equation*}
G(1+a) = \sum_{n=1}^\infty f^{(1)}_n
\left( \frac{1}{n+a+\varepsilon} + \frac{1}{n-a-2\varepsilon} \right)
+ \sum_{n=1}^\infty f^{(2)}_n
\left( \frac{1}{n+a} + \frac{1}{n-a-3\varepsilon} \right)
\end{equation*}
(where the symmetry~(\ref{Uni:sym}) is taken into account),
we obtain recurrence relations for the residues:
\begin{equation*}
f^{(1)}_n = - \frac{n+\varepsilon}{n+1-2\varepsilon} f^{(1)}_{n+1}\,,\quad
f^{(2)}_n = - \frac{n}{n+1-3\varepsilon} f^{(2)}_{n+1}\,.
\end{equation*}
Their solution is
\begin{equation*}
f^{(1)}_n = (-1)^n \frac{\Gamma(n+1-2\varepsilon)}{\Gamma(n+\varepsilon)} c_1(\varepsilon)\,,\quad
f^{(2)}_n = (-1)^n  \frac{\Gamma(n+1-3\varepsilon)}{\Gamma(n)} c_2(\varepsilon)\,,
\end{equation*}
where the constants are obtained from the initial condition:
\begin{equation*}
c_1(\varepsilon) = \frac{\Gamma(\varepsilon)}{\Gamma(2-2\varepsilon)}\,,\quad
c_2(\varepsilon) = -
\frac{\Gamma(\varepsilon) \Gamma(1-\varepsilon) \Gamma(1+\varepsilon)}%
{\Gamma(2-2\varepsilon) \Gamma(1-2\varepsilon) \Gamma(1+2\varepsilon)}\,.
\end{equation*}
Therefore we arrive at
\begin{align*}
&I(1+a) = 2
\frac{\Gamma^2(1-\varepsilon) \Gamma(\varepsilon) \Gamma(a+2\varepsilon) \Gamma(-a-\varepsilon)}%
{\Gamma(2-2\varepsilon) \Gamma(1+a) \Gamma(1-a-3\varepsilon)}\\
&{}\times\Biggl[ \sum_{n=1}^\infty (-1)^n \frac{\Gamma(n+1-2\varepsilon)}{\Gamma(n+\varepsilon)}
\left( \frac{1}{n+a+\varepsilon} + \frac{1}{n-a-2\varepsilon} \right)\\
&\qquad{} -
\frac{\Gamma(1-\varepsilon) \Gamma(1+\varepsilon)}%
{\Gamma(1-2\varepsilon) \Gamma(1+2\varepsilon)}
\sum_{n=1}^\infty (-1)^n \frac{\Gamma(n+1-3\varepsilon)}{\Gamma(n)}
\left( \frac{1}{n+a} + \frac{1}{n-a-3\varepsilon} \right)
\Biggr]\,.
\end{align*}
This result can be written via hypergeometric functions:
\begin{align}
&I(1+a) = 2
\frac{\Gamma(\varepsilon) \Gamma^2(1-\varepsilon) \Gamma(a+2\varepsilon) \Gamma(-a-\varepsilon)}%
{\Gamma(2-2\varepsilon)}
\Biggl\{
\frac{\Gamma(2-2\varepsilon)}%
{\Gamma(1+\varepsilon) \Gamma(1+a) \Gamma(1-a-3\varepsilon)}
\nonumber\\
&{}\times\Biggl[ \frac{1}{1+a+\varepsilon}
\F{3}{2}{1,2-2\varepsilon,1+a+\varepsilon\\1+\varepsilon,2+a+\varepsilon}{-1}
\nonumber\\
&\qquad{} + \frac{1}{1-a-2\varepsilon}
\F{3}{2}{1,2-2\varepsilon,1-a-2\varepsilon\\1+\varepsilon,2-a-2\varepsilon}{-1}
\Biggr]
- \cos(\pi\varepsilon) \Biggr\}\,.
\label{Uni:F32}
\end{align}

Kazakov~\cite{K:84,K:85} also derived several terms of expansion
of $I(a_i)$ with $a_i=1+n_i\varepsilon$ in $\varepsilon$
using symmetry properties of $I$;
we shall discuss this expansion in Sect.~\ref{S:Sym}.

\section{Symmetry}
\label{S:Sym}

Symmetries of the integrals~(\ref{Intro:mom})
which follow from inversion~(\ref{Intro:inv}),
duality between $x$ and $p$ space~(\ref{Intro:coord}),
and the star--triangle relation~(\ref{Uni:st})
were considered in~\cite{VPK:81}

\begin{figure}[ht]
\begin{center}
\begin{picture}(45,39)
\put(22.5,19.5){\makebox(0,0){\includegraphics[width=45mm]{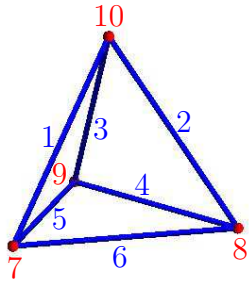}}}
\put(17.5,20){\makebox(0,0){\textcolor{blue}{1}}}
\put(35.5,22){\makebox(0,0){\textcolor{blue}{2}}}
\put(24.5,21){\makebox(0,0){\textcolor{blue}{3}}}
\put(30,13.5){\makebox(0,0){\textcolor{blue}{4}}}
\put(19,9){\makebox(0,0){\textcolor{blue}{5}}}
\put(27,4){\makebox(0,0){\textcolor{blue}{6}}}
\put(13,2.5){\makebox(0,0){\textcolor{red}{7}}}
\put(43,5){\makebox(0,0){\textcolor{red}{8}}}
\put(19,15){\makebox(0,0){\textcolor{red}{9}}}
\put(25.5,36){\makebox(0,0){\textcolor{red}{10}}}
\end{picture}
\end{center}
\caption{The tetrahedron diagram.}
\label{F:tetra}
\end{figure}

Gorishnii and Isaev~\cite{GI:84} discovered the tetrahedron symmetry group $S_4$
of the integrals $I$.
Let's consider the vacuum diagram in Fig.~\ref{F:tetra},
all lines have mass $m$.
If we integrate in the momentum of the line 6 last, then
\begin{equation*}
I = \frac{1}{\pi^{d/2}} \int
\frac{F(k^2)\,d^d k}{(k^2+m^2)^{a_6}}\,,
\end{equation*}
where $F(k^2)$ is the self-energy diagram with external momentum $k$
obtained by cutting the line 6.
Its asymptotics is
\begin{equation*}
F(k^2\to\infty) \to
\frac{I(a_1,a_2,a_3,a_4,a_5)}{(k^2)^{a_1+a_2+a_3+a_4+a_5-d}}
\end{equation*}
(it comes from the hard region;
other regions give contributions with different powers of $k^2$).
Hence the vacuum diagram has the ultraviolet pole
\begin{equation*}
I_{\text{UV}} = \frac{1}{\Gamma(d/2)}
\frac{I(a_1,a_2,a_3,a_4,a_5)}{a_1+a_2+a_3+a_4+a_5+a_6-\frac{3}{2}d}
\end{equation*}
(other regions produce poles at different places).
But we can equally well cut some other line,
$I_{\text{UV}}$ must remain intact.
Therefore $I(a_1,a_2,a_3,a_4,a_5)$ (which also depends on $d$)
can be considered as a function of $a_1,a_2,a_3,a_4,a_5,a_6$,
where $a_6$ is defined by
\begin{equation}
a_1 + a_2 + a_3 + a_4 + a_5 + a_6 = \frac{3}{2} d\,,
\label{Sym:a6}
\end{equation}
with the tetrahedron symmetry.

Gorishnii and Isaev also considered symmetry relations
following from the star--triangle relation~(\ref{Uni:st})
(which were discussed in~\cite{VPK:81,K:85}).
Taken together, these symmetry transformations are sufficient
for generating the complete symmetry group of the integrals $I$.
But they could not identify this group.
A complete solution of this problem was obtained in~\cite{B:86,BB:88}.

Let's introduce notation
\begin{equation}
I(a_1,a_2,a_3,a_4,a_5) =
\left[\prod_{i=1}^{10} G(a_i)\right]^{1/2}
\frac{\bar{I}(a_1,a_2,a_3,a_4,a_5,a_6)}%
{(d-3) \Gamma^2\left(\frac{d}{2}-1\right)}\,,
\label{Sym:barI}
\end{equation}
where $a_6$ is defined by~(\ref{Sym:a6}),
$\bar{a}_i$ by~(\ref{Intro:dual}),
indices of the vertices (Fig.~\ref{F:tetra}) are
\begin{equation*}
a_7 = a_1 + a_5 + a_6 - \frac{d}{2}\,,\quad
a_8 = a_2 + a_4 + a_6 - \frac{d}{2}\,,\quad
a_9 = a_3 + a_4 + a_5 - \frac{d}{2}\,,\quad
a_{10} = a_1 + a_2 + a_3 - \frac{d}{2}\,,
\end{equation*}
and
\begin{equation*}
G(a) = \frac{\Gamma(\bar{a})}{\Gamma(a)}\,.
\end{equation*}
The pre-factor in~(\ref{Sym:barI}) is chosen in such a way
that $\Gamma$-function factors
in~(\ref{Uni:comb}), (\ref{Uni:st}) cancel in symmetry relations.

The full symmetry group is generated by 3 transformations.
The first two generators are elements of the tetrahedron group:
\begin{align}
&\raisebox{-19mm}{\begin{picture}(45,39)
\put(22.5,19.5){\makebox(0,0){\includegraphics[width=45mm]{tetra.eps}}}
\put(17.5,20){\makebox(0,0){\textcolor{blue}{1}}}
\put(35.5,22){\makebox(0,0){\textcolor{blue}{2}}}
\put(24.5,21){\makebox(0,0){\textcolor{blue}{3}}}
\put(30,13.5){\makebox(0,0){\textcolor{blue}{4}}}
\put(19,9){\makebox(0,0){\textcolor{blue}{5}}}
\put(27,4){\makebox(0,0){\textcolor{blue}{6}}}
\put(13,2.5){\makebox(0,0){\textcolor{red}{7}}}
\put(43,5){\makebox(0,0){\textcolor{red}{8}}}
\put(19,15){\makebox(0,0){\textcolor{red}{9}}}
\put(25.5,36){\makebox(0,0){\textcolor{red}{10}}}
\end{picture}}
\hspace{5mm}\to
\raisebox{-19mm}{\begin{picture}(45,39)
\put(22.5,19.5){\makebox(0,0){\includegraphics[width=45mm]{tetra.eps}}}
\put(17.5,20){\makebox(0,0){\textcolor{blue}{3}}}
\put(35.5,22){\makebox(0,0){\textcolor{blue}{5}}}
\put(24.5,21){\makebox(0,0){\textcolor{blue}{4}}}
\put(30,13.5){\makebox(0,0){\textcolor{blue}{6}}}
\put(19,9){\makebox(0,0){\textcolor{blue}{2}}}
\put(27,4){\makebox(0,0){\textcolor{blue}{1}}}
\put(13,2.5){\makebox(0,0){\textcolor{red}{10}}}
\put(43,5){\makebox(0,0){\textcolor{red}{7}}}
\put(19,15){\makebox(0,0){\textcolor{red}{8}}}
\put(25.5,36){\makebox(0,0){\textcolor{red}{9}}}
\end{picture}}\,,
\label{Sym:gen1}\\
&\raisebox{-19mm}{\begin{picture}(45,39)
\put(22.5,19.5){\makebox(0,0){\includegraphics[width=45mm]{tetra.eps}}}
\put(17.5,20){\makebox(0,0){\textcolor{blue}{1}}}
\put(35.5,22){\makebox(0,0){\textcolor{blue}{2}}}
\put(24.5,21){\makebox(0,0){\textcolor{blue}{3}}}
\put(30,13.5){\makebox(0,0){\textcolor{blue}{4}}}
\put(19,9){\makebox(0,0){\textcolor{blue}{5}}}
\put(27,4){\makebox(0,0){\textcolor{blue}{6}}}
\put(13,2.5){\makebox(0,0){\textcolor{red}{7}}}
\put(43,5){\makebox(0,0){\textcolor{red}{8}}}
\put(19,15){\makebox(0,0){\textcolor{red}{9}}}
\put(25.5,36){\makebox(0,0){\textcolor{red}{10}}}
\end{picture}}
\hspace{5mm}\to
\raisebox{-19mm}{\begin{picture}(45,39)
\put(22.5,19.5){\makebox(0,0){\includegraphics[width=45mm]{tetra.eps}}}
\put(17.5,20){\makebox(0,0){\textcolor{blue}{2}}}
\put(35.5,22){\makebox(0,0){\textcolor{blue}{1}}}
\put(24.5,21){\makebox(0,0){\textcolor{blue}{3}}}
\put(30,13.5){\makebox(0,0){\textcolor{blue}{5}}}
\put(19,9){\makebox(0,0){\textcolor{blue}{4}}}
\put(27,4){\makebox(0,0){\textcolor{blue}{6}}}
\put(13,2.5){\makebox(0,0){\textcolor{red}{8}}}
\put(43,5){\makebox(0,0){\textcolor{red}{7}}}
\put(19,15){\makebox(0,0){\textcolor{red}{9}}}
\put(25.5,36){\makebox(0,0){\textcolor{red}{10}}}
\end{picture}}\,.
\label{Sym:gen2}
\end{align}
The last one comes from uniqueness.
First we introduce an extra dot on line 3 to make vertex 10 unique;
then use the star--triangle relation~(\ref{Uni:st});
and then combine two lines:
\begin{align*}
&\raisebox{-15mm}{\begin{picture}(26,31)
\put(13,14.25){\makebox(0,0){\includegraphics{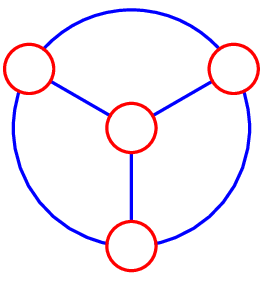}}}
\put(7,16.5){\makebox(0,0){\textcolor{blue}{1}}}
\put(19,16.5){\makebox(0,0){\textcolor{blue}{2}}}
\put(13.5,9.5){\makebox(0,0)[l]{\textcolor{blue}{3}}}
\put(25.124,8.5){\makebox(0,0){\textcolor{blue}{4}}}
\put(0.876,8.5){\makebox(0,0){\textcolor{blue}{5}}}
\put(13,29.5){\makebox(0,0){\textcolor{blue}{6}}}
\put(2.608,21.5){\makebox(0,0){\textcolor{red}{7}}}
\put(23.392,21.5){\makebox(0,0){\textcolor{red}{8}}}
\put(13,3.5){\makebox(0,0){\textcolor{red}{9}}}
\put(13,15.5){\makebox(0,0){\textcolor{red}{10}}}
\end{picture}}
\quad\to\quad
\raisebox{-15mm}{\begin{picture}(26,31)
\put(13,14.25){\makebox(0,0){\includegraphics{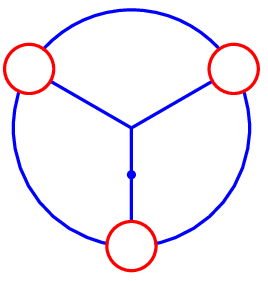}}}
\put(7,16.5){\makebox(0,0){\textcolor{blue}{1}}}
\put(19,16.5){\makebox(0,0){\textcolor{blue}{2}}}
\put(13.5,13.125){\makebox(0,0)[l]{\textcolor{blue}{$\overline{1}+\overline{2}$}}}
\put(13.5,8.375){\makebox(0,0)[l]{\textcolor{blue}{10}}}
\put(25.124,8.5){\makebox(0,0){\textcolor{blue}{4}}}
\put(0.876,8.5){\makebox(0,0){\textcolor{blue}{5}}}
\put(13,29.5){\makebox(0,0){\textcolor{blue}{6}}}
\put(2.608,21.5){\makebox(0,0){\textcolor{red}{7}}}
\put(23.392,21.5){\makebox(0,0){\textcolor{red}{8}}}
\put(13,3.5){\makebox(0,0){\textcolor{red}{$\overline{6}$}}}
\end{picture}}
\quad\to{}\displaybreak\\
&\raisebox{-15mm}{\begin{picture}(26,31)
\put(13,14.25){\makebox(0,0){\includegraphics{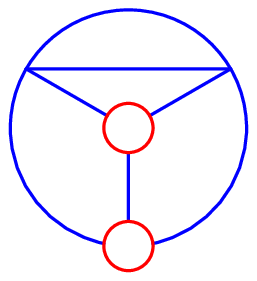}}}
\put(7,16.5){\makebox(0,0){\textcolor{blue}{$\overline{2}$}}}
\put(19,16.5){\makebox(0,0){\textcolor{blue}{$\overline{1}$}}}
\put(13.5,9.5){\makebox(0,0)[l]{\textcolor{blue}{10}}}
\put(25.124,8.5){\makebox(0,0){\textcolor{blue}{4}}}
\put(0.876,8.5){\makebox(0,0){\textcolor{blue}{5}}}
\put(13,29.5){\makebox(0,0){\textcolor{blue}{6}}}
\put(13,3.5){\makebox(0,0){\textcolor{red}{$\overline{6}$}}}
\put(13,15.5){\makebox(0,0){\textcolor{red}{3}}}
\put(13,22){\makebox(0,0)[b]{\textcolor{blue}{$1+2$}}}
\end{picture}}
\quad\to\quad
\raisebox{-15mm}{\begin{picture}(26,31)
\put(13,14.25){\makebox(0,0){\includegraphics{u1.eps}}}
\put(7,16.5){\makebox(0,0){\textcolor{blue}{$\overline{2}$}}}
\put(19,16.5){\makebox(0,0){\textcolor{blue}{$\overline{1}$}}}
\put(13.5,9.5){\makebox(0,0)[l]{\textcolor{blue}{10}}}
\put(25.124,8.5){\makebox(0,0){\textcolor{blue}{4}}}
\put(0.876,8.5){\makebox(0,0){\textcolor{blue}{5}}}
\put(13,29.5){\makebox(0,0){\textcolor{blue}{9}}}
\put(2.608,21.5){\makebox(0,0){\textcolor{red}{7}}}
\put(23.392,21.5){\makebox(0,0){\textcolor{red}{8}}}
\put(13,3.5){\makebox(0,0){\textcolor{red}{$\overline{6}$}}}
\put(13,15.5){\makebox(0,0){\textcolor{red}{3}}}
\end{picture}}\quad.
\end{align*}

I.\,e., the third generator is
\begin{equation}
\raisebox{-19mm}{\begin{picture}(45,39)
\put(22.5,19.5){\makebox(0,0){\includegraphics[width=45mm]{tetra.eps}}}
\put(17.5,20){\makebox(0,0){\textcolor{blue}{1}}}
\put(35.5,22){\makebox(0,0){\textcolor{blue}{2}}}
\put(24.5,21){\makebox(0,0){\textcolor{blue}{3}}}
\put(30,13.5){\makebox(0,0){\textcolor{blue}{4}}}
\put(19,9){\makebox(0,0){\textcolor{blue}{5}}}
\put(27,4){\makebox(0,0){\textcolor{blue}{6}}}
\put(13,2.5){\makebox(0,0){\textcolor{red}{7}}}
\put(43,5){\makebox(0,0){\textcolor{red}{8}}}
\put(19,15){\makebox(0,0){\textcolor{red}{9}}}
\put(25.5,36){\makebox(0,0){\textcolor{red}{10}}}
\end{picture}}
\hspace{5mm}\to
\raisebox{-19mm}{\begin{picture}(45,39)
\put(22.5,19.5){\makebox(0,0){\includegraphics[width=45mm]{tetra.eps}}}
\put(17.5,20){\makebox(0,0){\textcolor{blue}{$\overline{2}$}}}
\put(35.5,22){\makebox(0,0){\textcolor{blue}{$\overline{1}$}}}
\put(24.5,21){\makebox(0,0){\textcolor{blue}{10}}}
\put(30,13.5){\makebox(0,0){\textcolor{blue}{4}}}
\put(19,9){\makebox(0,0){\textcolor{blue}{5}}}
\put(27,4){\makebox(0,0){\textcolor{blue}{$\overline{9}$}}}
\put(13,2.5){\makebox(0,0){\textcolor{red}{7}}}
\put(43,5){\makebox(0,0){\textcolor{red}{8}}}
\put(19,15){\makebox(0,0){\textcolor{red}{$\overline{6}$}}}
\put(25.5,36){\makebox(0,0){\textcolor{red}{3}}}
\end{picture}}\quad.
\label{Sym:gen3}
\end{equation}

The structure of the group becomes apparent if we introduce new variables:
\begin{align}
&\bar{I}(a_1,a_2,a_3,a_4,a_5,a_6) = \tilde{I}(b_1,b_2,b_3,b_4,b_5,b_6)\,,
\label{Sym:b}\\
&\left(\begin{array}{c}
b_1\\b_2\\b_3\\b_4\\b_5\\b_6
\end{array}\right) =
\frac{1}{3}
\left(\begin{array}{rrrrrr}
  1 &  2 &  0 &  1 & -1 &  0 \\
  0 &  1 &  2 &  0 &  1 & -1 \\
 -1 &  0 &  1 &  2 &  0 &  1 \\
  1 & -1 &  0 &  1 &  2 &  0 \\
  0 &  1 & -1 &  0 &  1 &  2 \\
  2 &  0 &  1 & -1 &  0 &  1
\end{array}\right)
\left(\begin{array}{c}
a_1\\a_2\\a_3\\a_4\\a_5\\a_6
\end{array}\right)\,.
\nonumber
\end{align}
Then our 3 generators transform $\tilde{I}(b_1,b_2,b_3,b_4,b_5,b_6)$ to
\begin{align*}
&\hat{P}_1 \tilde{I}(b_1,b_2,b_3,b_4,b_5,b_6) =
\tilde{I}(\bar{b}_1,\bar{b}_6,\bar{b}_2,\bar{b}_4,\bar{b}_3,\bar{b}_5)\,,\\
&\hat{P}_2 \tilde{I}(b_1,b_2,b_3,b_4,b_5,b_6) =
\tilde{I}(\bar{b}_3,\bar{b}_5,\bar{b}_1,\bar{b}_6,\bar{b}_2,\bar{b}_4)\,,\\
&\hat{P}_3 \tilde{I}(b_1,b_2,b_3,b_4,b_5,b_6) =
\tilde{I}(b_3,b_2,b_1,b_4,b_5,b_6)\,.
\end{align*}
We can combine them into 3 better generators
\begin{equation*}
\hat{Q}_1 = (\hat{P}_3 \hat{P}_1^2)^2 \hat{P}_3 \hat{P}_2 \hat{P}_1\,,\quad
\hat{Q}_2 = \hat{P}_1^3 \hat{P}_3 \hat{P}_1\,,\quad
\hat{Q}_3 = (\hat{P}_3 \hat{P}_2 \hat{P}_1^2)^2\,;
\end{equation*}
they transform $\tilde{I}(b_1,b_2,b_3,b_4,b_5,b_6)$ to
\begin{align*}
&\hat{Q}_1 \tilde{I}(b_1,b_2,b_3,b_4,b_5,b_6) =
\tilde{I}(b_2,b_3,b_4,b_5,b_6,b_1)\,,\\
&\hat{Q}_2 \tilde{I}(b_1,b_2,b_3,b_4,b_5,b_6) =
\tilde{I}(b_2,b_1,b_3,b_4,b_5,b_6)\,,\\
&\hat{Q}_3 \tilde{I}(b_1,b_2,b_3,b_4,b_5,b_6) =
\tilde{I}(\bar{b}_1,\bar{b}_2,\bar{b}_3,\bar{b}_4,\bar{b}_5,\bar{b}_6)\,.
\end{align*}
The first two generate the symmetric group $S_6$
of permutations of 6 variables $b_{1,\ldots,6}$;
and $\hat{Q}_3$ generates $Z_2$.
So, the symmetry group of the integrals $I$ is $S_6\times Z_2$~\cite{BB:88};
it contains $6!\cdot2=1440$ elements~\cite{B:86}.

The most useful information is the expansion
of $I(a_1,a_2,a_3,a_4,a_5)$~(\ref{Intro:mom}) in $\varepsilon$
at $d=4-2\varepsilon$ and $a_{1,\ldots,5}=1+\mathcal{O}(\varepsilon)$.
This means that all $a_i$ are $1+\mathcal{O}(\varepsilon)$;
the same is true for $b_i$
(note that $\sum_{i=1}^6 b_i = \frac{3}{2} d$).
The function~(\ref{Sym:b}) is invariant with respect to $S_6\times Z_2$;
therefore, the expansion can be written entirely via invariants of this group.
The invariants are
\begin{equation*}
I_1 = 1 - \frac{d}{4} = 1 - \frac{1}{6} \sum_{i=1}^6 b_i
\end{equation*}
and
\begin{equation*}
I_n = \sum_{i=1}^6 \left(b_i - \frac{d}{4}\right)^n
\end{equation*}
for $n=2$, 3, 4, 5, 6.
The total degree of $I_3$ and $I_5$ must be even,
because they change their signs under $\hat{Q}_5$:
\begin{equation}
\bar{I}(a_1,a_2,a_3,a_4,a_5,a_6) =
\sum_{i_3+i_5\;\text{even}} C_{i_1 i_2 i_3 i_4 i_5 i_6}
I_1^{i_1} I_2^{i_2} I_3^{i_3} I_4^{i_4} I_5^{i_5} I_6^{i_6}\,.
\label{Sym:exp}
\end{equation}
All unknown coefficients $C_{i_1 i_2 i_3 i_4 i_5 i_6}$
needed for expanding up to $\varepsilon^4$
can be fixed by considering the integrals~(\ref{IBP:red})
(the indices in the left triangle are 1).
IBP reduces these integrals to $\Gamma$-functions,
and hence all these coefficients are expressed via $\zeta_n$.
So, the symmetry allows one to obtain,
practically for free~\cite{K:84,K:85,B:86,BB:88},
\begin{align}
&\bar{I}(a_1,a_2,a_3,a_4,a_5,a_6) =
6 \zeta_3 + 18 \zeta_4 I_1 + 3 \zeta_5 \left(I_1^2 + \frac{5}{2} I_2\right)
\label{Sym:e4}\\
&{} - 15 (7 \zeta_6 + 2 \zeta_3^2) I_1^3
+ 3 \left(\frac{25}{2} \zeta_6 - \zeta_3^2\right) I_1 I_2
\nonumber\\
&{} - 9 \left(\frac{439}{8} \zeta_7 + 20 \zeta_4 \zeta_3\right) I_1^4
+ 3 \left(\frac{211}{8} \zeta_7 - 6 \zeta_4 \zeta_3\right) I_1^2 I_2
+ \frac{9}{8} \zeta_7 \left(\frac{35}{4} I_2^2 - 7 I_4\right)
+ \cdots
\nonumber
\end{align}

\section{Gegenbauer polynomials}
\label{S:Geg}

\begin{figure}[ht]
\begin{center}
\begin{picture}(98,24)
\put(15,12){\makebox(0,0){\includegraphics{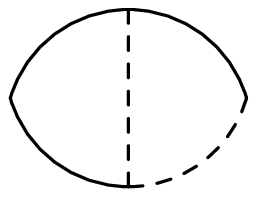}}}
\put(1.5,12){\makebox(0,0){0}}
\put(28.5,12){\makebox(0,0){$z$}}
\put(15,22.5){\makebox(0,0){$x$}}
\put(15,1.5){\makebox(0,0){$y$}}
\put(6,4){\makebox(0,0){$a$}}
\put(24,20){\makebox(0,0){$b$}}
\put(6,20){\makebox(0,0){$c$}}
\put(52,12){\makebox(0,0){\includegraphics{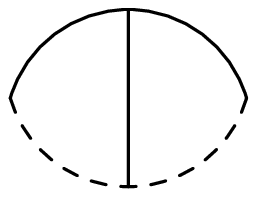}}}
\put(86,12){\makebox(0,0){\includegraphics{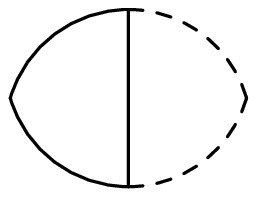}}}
\end{picture}
\end{center}
\caption{Integrals with 2 adjacent lines having indices 1.}
\label{F:Kotikov}
\end{figure}

If indices of 3 lines forming a triangle are 1,
$I$ reduces to $\Gamma$-functions by IBP~(\ref{IBP:red}).
In a more general case when 2 adjacent lines have indices 1
$I$ can be expressed via hypergeometric functions of 1.
Due to the tetrahedron symmetry, it does not matter
which 2 adjacent lines have indices 1 (Fig.~\ref{F:Kotikov}).
These integrals were calculated by Kotikov~\cite{K:96}
by an ingenious use of $x$-space Gegenbauer polynomials~\cite{CKT:80}.

Let's consider the first of these diagrams in $x$-space:
\begin{align}
&I(c,b,1,1,a) =
\frac{G^2(1) G(a) G(b) G(c)}{G(a+b+c+2-d)}
A(\bar{a},\bar{b},\bar{c})\,,
\label{Geg:A}\\
&A(a,b,c) = \frac{1}{\pi^d} \int \frac{d^d x\,d^d y}%
{(y^2)^a [(z-y)^2]^\lambda [(z-x)^2]^b (x^2)^c [(x-y)^2]^\lambda}\,,
\nonumber
\end{align}
where $\lambda=d/2-1$ and $z^2=1$.
Expanding the propagator as
\begin{equation}
\frac{1}{[(y-z)^2]^\lambda} = \frac{1}{\Gamma(\lambda)}
\sum_{n=0}^\infty \frac{\Gamma(\lambda+n)}{n!}
y^{\mu_1\ldots\mu_n} z_{\mu_1\ldots\mu_n}
\left[ \theta(1-y^2) + \frac{\theta(y^2-1)}{(y^2)^{n+\lambda}} \right]
\label{Geg:exp}
\end{equation}
(where $y^{\mu_1\ldots\mu_n}$ is the traceless part of $y^{\mu_1}\cdots y^{\mu_n}$)
we can integrate in $d^d y$:
\begin{align*}
&A(a,b,c) = \frac{1}{\Gamma^2(\lambda) (a-1)}
\sum_{n=0}^\infty \frac{2^n \Gamma(n+\lambda)}{n!} \frac{z_{\mu_1\ldots\mu_n}}{\pi^{d/2}}
\int \frac{d^d x\,x^{\mu_1\ldots\mu_n}}{(x^2)^c [(z-x)^2]^b}\\
&\left[ \frac{1}{n+\lambda-a+1}
\left( \frac{\theta(1-x^2)}{(x^2)^{a-1}} + \frac{\theta(x^2-1)}{(x^2)^{n+\lambda}} \right)
- \frac{1}{n+\lambda+a-1}
\left( \theta(1-x^2) + \frac{\theta(x^2-1)}{(x^2)^{n+\lambda+a-1}} \right) \right]\,.
\end{align*}
Now the integral in $d^d x$ can be calculated:
\begin{align*}
&A(a,b,c) = \frac{1}{\Gamma(\lambda) \Gamma(2\lambda) \Gamma(b) \Gamma(b-\lambda) (a-1)}
\sum_{n=0}^\infty \frac{2^n \Gamma(n+\lambda)}{n!}
\sum_{m=0}^\infty \frac{\Gamma(m+n+b) \Gamma(m+b-\lambda)}{m! \Gamma(m+n+\lambda+1)}\\
&\biggl[  \frac{1}{n+\lambda-a+1}
\left( \frac{1}{m+n-a-c+\lambda+2} + \frac{1}{m+n+b+c-1} \right)\\
&\quad{} - \frac{1}{n+\lambda+a-1}
\left( \frac{1}{m+n-c+\lambda+1} + \frac{1}{m+n+a+b+c-2} \right) \biggr]\,.
\end{align*}

It appears to be possible to transform this result
into a form containing only single sums~\cite{K:96}:
\begin{equation}
A(a,b,c) = \frac{A_1 - A_2}{\Gamma\left(\frac{d}{2}-1\right) (a-1)}
\label{Geg:res}
\end{equation}
where
\begin{align}
&A_1 = 2 \frac{\Gamma\left(\frac{d}{2}-b\right)}{\Gamma(b)}
\biggl\{ \frac{1}{d-2a}
\label{Geg:A1}\\
&\biggl[
\frac{\Gamma\left(\frac{d}{2}-a-c+1\right) \Gamma\left(a+b+c-\frac{d}{2}-1\right)}%
{\Gamma(a+c-1) \Gamma(d-a-b-c+1)}
\F{3}{2}{d-2,\frac{d}{2}-a,\frac{d}{2}-a-c+1\\\frac{d}{2}-a+1,d-a-b-c+1}{1}
\nonumber\\
&\quad{} +
\frac{\Gamma(1-c) \Gamma(b+c-1)}%
{\Gamma\left(c+\frac{d}{2}-1\right) \Gamma\left(\frac{d}{2}-b-c+1\right)}
\F{3}{2}{d-2,\frac{d}{2}-a,b+c-1\\\frac{d}{2}-a+1,c+\frac{d}{2}-1}{1}
\biggr]
\nonumber\\
&{} - \frac{1}{2a+d-4} \biggl[
\frac{\Gamma\left(\frac{d}{2}-c\right) \Gamma\left(b+c-\frac{d}{2}\right)}%
{\Gamma(c) \Gamma(d-b-c)}
\F{3}{2}{d-2,a+\frac{d}{2}-2,\frac{d}{2}-c\\a+\frac{d}{2}-1,d-b-c}{1}
\nonumber\displaybreak\\
&{} +
\frac{\Gamma(2-a-c) \Gamma(a+b+c-2)}%
{\Gamma\left(a+c+\frac{d}{2}-2\right) \Gamma\left(\frac{d}{2}-a-b-c+2\right)}
\F{3}{2}{d-2,a+\frac{d}{2}-2,a+b+c-2\\a+\frac{d}{2}-1,a+c+\frac{d}{2}-2}{1}
\biggr] \biggr\}\,,
\nonumber\\
&A_2 =
\frac{\Gamma(1-b) \Gamma(1-c) \Gamma\left(\frac{d}{2}-a\right) \Gamma\left(a+\frac{d}{2}-2\right)
\Gamma\left(\frac{d}{2}-b\right) \Gamma\left(a+b+c-\frac{d}{2}-1\right)}%
{\Gamma(d-2) \Gamma(a+c-1) \Gamma\left(a+b-\frac{d}{2}\right)
\Gamma\left(\frac{d}{2}-a-b+1\right) \Gamma\left(\frac{d}{2}-b-c+1\right)}
\nonumber\\
&{} - \frac{2 \Gamma(1-b)}{(2a+d-4) \Gamma\left(b-\frac{d}{2}+1\right)}
\nonumber\\
&\biggl[
\frac{\Gamma(2-a-c) \Gamma\left(a+b+c-\frac{d}{2}-1\right)}%
{\Gamma(3-a-b-c) \Gamma\left(a+c+\frac{d}{2}-2\right)}
\F{3}{2}{d-2,a+\frac{d}{2}-2,a+b+c-2\\a+\frac{d}{2}-1,a+c+\frac{d}{2}-2}{1}
\nonumber\\
&\quad{} +
\frac{\Gamma(1-c) \Gamma\left(b+c-\frac{d}{2}\right)}%
{\Gamma\left(c-\frac{d}{2}+1\right) \Gamma(d-b-c)}
\F{3}{2}{d-2,a+\frac{d}{2}-2,\frac{d}{2}-c\\a+\frac{d}{2}-1,d-b-c}{1}
\biggr]
\label{Geg:A21}\\
&{} = -
\frac{\Gamma(1-b) \Gamma(2-a-c) \Gamma\left(\frac{d}{2}-a\right) \Gamma\left(a+\frac{d}{2}-2\right)
\Gamma\left(\frac{d}{2}-b\right) \Gamma\left(b+c-\frac{d}{2}\right)}%
{\Gamma(c) \Gamma(d-2) \Gamma\left(b-a-\frac{d}{2}+2\right) \Gamma\left(a-b+\frac{d}{2}-1\right)
\Gamma\left(\frac{d}{2}-a-b-c+2\right)}
\nonumber\\
&{} + \frac{2 \Gamma(1-b)}{(d-2a) \Gamma\left(b-\frac{d}{2}+1\right)}
\biggl[
\frac{\Gamma(1-c) \Gamma\left(b+c-\frac{d}{2}\right)}%
{\Gamma(2-b-c) \Gamma\left(c+\frac{d}{2}-1\right)}
\F{3}{2}{d-2,\frac{d}{2}-a,b+c-1\\\frac{d}{2}-a+1,c+\frac{d}{2}-1}{1}
\nonumber\\
&{} +
\frac{\Gamma(2-a-c) \Gamma\left(a+b+c-\frac{d}{2}-1\right)}%
{\Gamma\left(a+c-\frac{d}{2}\right) \Gamma(d-a-b-c+1)}
\F{3}{2}{d-2,\frac{d}{2}-a,\frac{d}{2}-a-c+1\\\frac{d}{2}-a+1,d-a-b-c+1}{1}
\biggr]
\label{Geg:A22}\\
&{} =
\frac{\Gamma(1-b) \Gamma(1-c) \Gamma(2-a-c) \Gamma\left(\frac{d}{2}-a\right)
\Gamma\left(a+\frac{d}{2}-2\right) \Gamma\left(\frac{d}{2}-b\right)}%
{\Gamma(1-a) \Gamma(a) \Gamma(d-2) \Gamma\left(\frac{d}{2}-a-b+1\right)
\Gamma\left(\frac{d}{2}-a-b-c+2\right)}
\nonumber\\
&{} + \frac{2 \Gamma(1-b)}{\Gamma\left(b-\frac{d}{2}+1\right)}
\biggl[
\frac{\Gamma(2-a-c) \Gamma\left(a+b+c-\frac{d}{2}-1\right)}%
{(d-2a) \Gamma\left(a+c-\frac{d}{2}\right) \Gamma(d-a-b-c+1)}
\nonumber\\
&\qquad{}
\F{3}{2}{d-2,\frac{d}{2}-a,\frac{d}{2}-a-c+1\\\frac{d}{2}-a+1,d-a-b-c+1}{1}
\nonumber\\
&{} -
\frac{\Gamma(1-c) \Gamma\left(b+c-\frac{d}{2}\right)}%
{(2a+d-4) \Gamma\left(c-\frac{d}{2}+1\right) \Gamma(d-b-c)}
\F{3}{2}{d-2,a+\frac{d}{2}-2,\frac{d}{2}-c\\a+\frac{d}{2}-1,d-b-c}{1}
\biggr]
\label{Geg:A23}\\
&{} =
\frac{\Gamma(1-b) \Gamma\left(\frac{d}{2}-a\right) \Gamma\left(a+\frac{d}{2}-2\right)
\Gamma\left(\frac{d}{2}-b\right) \Gamma\left(b+c-\frac{d}{2}\right)
\Gamma\left(a+b+c-\frac{d}{2}-1\right)}%
{\Gamma(1-a) \Gamma(a) \Gamma(c) \Gamma(d-2) \Gamma\left(c+\frac{d}{2}-2\right)}
\nonumber\\
&{} + \frac{2 \Gamma(1-b)}{\Gamma \left(b-\frac{d}{2}+1\right)}
\biggl[
\frac{\Gamma(1-c) \Gamma\left(b+c-\frac{d}{2}\right)}%
{(d-2a) \Gamma(2-b-c) \Gamma\left(c+\frac{d}{2}-1\right)}
\F{3}{2}{d-2,\frac{d}{2}-a,b+c-1\\\frac{d}{2}-a+1,c+\frac{d}{2}-1}{1}
\nonumber\\
&{} -
\frac{\Gamma(2-a-c) \Gamma\left(a+b+c-\frac{d}{2}-1\right)}%
{(2a+d-4) \Gamma(3-a-b-c) \Gamma\left(a+c+\frac{d}{2}-2\right)}
\nonumber\\
&\qquad{}
\F{3}{2}{d-2,a+\frac{d}{2}-2,a+b+c-2\\a+\frac{d}{2}-1,a+c+\frac{d}{2}-2}{1}
\biggr]\,.
\label{Geg:A24}
\end{align}

In particular, for the integral $I(a)$~(\ref{Uni:Ia}) we obtain
\begin{align}
&I(a) = 2
\Gamma\left(\tfrac{d}{2}-1\right) \Gamma\left(\tfrac{d}{2}-a-1\right)
\Gamma(a-d+3)
\label{Geg:Ia}\\
&\biggl[
\frac{2\Gamma\left(\frac{d}{2}-1\right)}%
{(d-2a-4)\Gamma(a+1)\Gamma\left(\frac{3}{2}d-a-4\right)}
\F{3}{2}{1,d-2,a-\frac{d}{2}+2\\a+1,a-\frac{d}{2}+3}{1}
- \frac{\pi\cot\pi(d-a)}{\Gamma(d-2)} \biggr]\,;
\nonumber
\end{align}
this form is equivalent to~(\ref{Uni:F32}),
though mathematical proof is unknown.

No expression for $I$ with unit indices of 2 non-adjacent lines
is known.

\section{Solving IBP for 3 non-integer indices}
\label{S:Beyond}

Expressions for $I$ with 2 indices of adjacent lines equal to 1
via hypergeometric functions of 1 were also derived by guessing
the solution of IBP for such integrals~\cite{BGK:97}.
Let's consider
\begin{equation}
I(a_1,a_2,a_3,a_4) =
\raisebox{-7.5mm}{\begin{picture}(30,17)
\put(15,8.5){\makebox(0,0){\includegraphics{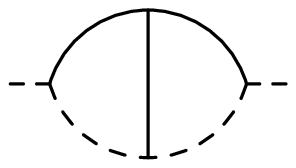}}}
\put(15.5,8){\makebox(0,0)[l]{$a_3$}}
\put(7,15){\makebox(0,0){$a_1$}}
\put(23,15){\makebox(0,0){$a_2$}}
\end{picture}}
\label{Beyond:I}
\end{equation}
where $a_4 = a_1 + a_2 + a_3 - \frac{d}{2}$.
The IBP relations are
\begin{align}
&a_1 I(a_1+1,a_2,a_3,a_4+1) - (a_1+a_2-d+2) I(a_1,a_2,a_3,a_4)
\nonumber\\
&{} = a_3 G(1,a_4+1)
\left( \frac{a_1 G(a_1+1,a_3+1)}{a_4-a_2+1} - G(a_2,a_3+1) \right)\,,
\nonumber\\
&(a_3-d+2) I(a_1,a_2,a_3,a_4)
\nonumber\\
&\quad{} + \frac{(a_1+a_3-d+1) (a_2+a_3-d+1)}{a_3-d/2+1} I(a_1,a_2,a_3-1,a_4-1)
\nonumber\\
&{} = a_3 (a_3+a_4-d+1) G(1,a_4)
\left( \frac{G(a_1,a_3+1)}{a_4-a_2} + \frac{G(a_2,a_3+1)}{a_4-a_1} \right)\,,
\label{Beyond:IBP}
\end{align}
where $G(a_1,a_2)=G(a_1) G(a_2) G(\bar{a_1}+\bar{a_2})$
is the standard massless one-loop self-energy.

If we express $I(a_1,a_2,a_3,a_4)$ as
\begin{align}
&\frac{d-3}{a_3 a_4 G(1,a_4+1)} I(a_1,a_2,a_3,a_4)
\nonumber\\
&{} = G(a_1,a_2+1)
S({\textstyle\frac{d}{2}}-a_1-1,a_2-1,{\textstyle\frac{d}{2}}+a_1-a_4-2,a_4-a_2)
+ (a_1\leftrightarrow a_2)
\label{Beyond:Ansatz}
\end{align}
via a new function $S(a_1,a_2,a_3,a_4)$ satisfying
\begin{align}
&S(a_1,a_2,a_3,a_4) = S(a_2,a_1,a_3,a_4) = - S(a_3,a_4,a_1,a_2)\,,
\nonumber\\
& a_1 S(a_1,a_2,a_3,a_4) = 1 + \frac{(a_1+a_3) (a_1+a_4)}{a_1+a_2+a_3+a_4} S(a_1-1,a_2,a_3,a_4)\,,
\label{Beyond:Sprop}
\end{align}
then~(\ref{Beyond:IBP}) holds.
The solution of~(\ref{Beyond:Sprop}) can be written as
\begin{equation}
S(a_1,a_2,a_3,a_4) =
\frac{\pi \cot \pi a_3}{H(a_1,a_2,a_3,a_4)} - \frac{1}{a_3}
- \frac{a_2+a_3}{a_2 a_3} F(a_1+a_3,-a_2,-a_3,a_2+a_4)
\label{Beyond:S}
\end{equation}
where
\begin{equation*}
H(a_1,a_2,a_3,a_4) =
\frac{\Gamma(1+a_1) \Gamma(1+a_2) \Gamma(1+a_3) \Gamma(1+a_4) \Gamma(1+a_1+a_2+a_3+a_4)}%
{\Gamma(1+a_1+a_3) \Gamma(1+a_1+a_4) \Gamma(1+a_2+a_3) \Gamma(1+a_2+a_4)}
\end{equation*}
and
\begin{equation}
F(a_1,a_2,a_3,a_4) = \F{3}{2}{1,-a_1,-a_2\\1+a_3,1+a_4}{1} - 1\,.
\label{Beyond:F32}
\end{equation}

Expansion of
\begin{equation*}
\raisebox{-4mm}{\begin{picture}(30,17)
\put(15,8.5){\makebox(0,0){\includegraphics{i4.eps}}}
\put(15.5,8.5){\makebox(0,0)[l]{$1+b$}}
\put(4.5,15){\makebox(0,0){$1+a$}}
\put(25.5,15){\makebox(0,0){$1+a$}}
\end{picture}}
\end{equation*}
to all orders in $a$, $b$ at $\varepsilon=0$
can be expressed via $\zeta_{2n+1}$~\cite{BGK:97}
(some particular cases were known earlier~\cite{K:85,B:86,BB:88}).

Elegant symmetry-based methods to derive several terms of $\varepsilon$ expansion
of hypergeometric functions~(\ref{Beyond:F32}) algebraically
are presented in~\cite{BGK:97}.
However, these methods cannot be extended to higher orders.
They are no longer necessary: the algorithm constructed in~\cite{MUW:02}
allow one to expand such hypergeometric functions to any order,
results are expressible via multiple $\zeta$ values.
The knowledge of expansions of these ${}_3F_2(1)$ allows one to reconstruct
all unknown coefficients in the general expansion~(\ref{Sym:exp})
up to $\varepsilon^9$~\cite{BGK:97,B:03}.
The $\varepsilon^5$ term
\begin{align}
&\bar{I}(a_1,a_2,a_3,a_4,a_5,a_6) = \cdots
\nonumber\\
&{} + 3 \left( \frac{378}{5} \zeta_{5\,3} - \frac{33523}{40} \zeta_8 + 10 \zeta_5 \zeta_3 \right) I_1^5
- 3 \left( \frac{54}{5} \zeta_{5\,3} - \frac{1009}{40} \zeta_8 + 42 \zeta_5 \zeta_3 \right) I_1^3 I_2
\nonumber\\
&\quad{} - \frac{3}{2} \left( \frac{9}{5} \zeta_{5\,3} - \frac{4023}{80} \zeta_8 + 7 \zeta_5 \zeta_3 \right) I_1 I_2^2
+ 3 \left( \frac{18}{5} \zeta_{5\,3} - \frac{1083}{40} \zeta_8 + 8 \zeta_5 \zeta_3 \right) I_1 I_4
+ \cdots
\label{Beyond:e5}
\end{align}
is the first term where a depth-2 value $\zeta_{5\,3}$ appears.

In particular, we obtain several equivalent results for $I(a)$:
\begin{align}
&\frac{(d-3)(d-4)\Gamma(a) \Gamma\left(\frac{3}{2}d-a-4\right)}%
{2 \Gamma^2\left(\frac{d}{2}-1\right) \Gamma(a-d+3) \Gamma\left(\frac{d}{2}-a-1\right)}
I(a)
\nonumber\\
&{} = \frac{3d-2a-10}{d-a-3}
\F{3}{2}{1,\frac{d}{2}-2,a-d+3\\a,a-d+4}{1}
+ A \pi \cot\pi(a-d) - 2
\label{Beyond:Ia1}\\
&{} = - \frac{3d-2a-10}{d-a-3}
\F{3}{2}{1,1-a,d-a-3\\3-\frac{d}{2},d-a-2}{1}
+ A \pi \cot\pi\tfrac{d}{2} + \frac{d-4}{d-a-3}
\label{Beyond:Ia2}\\
&{} = 4 \frac{a-1}{d-2a-2}
\F{3}{2}{1,a-\frac{3}{2}d+5,a-\frac{d}{2}+1\\3-\frac{d}{2},a-\frac{d}{2}+2}{1}
+ A \pi \cot\pi\tfrac{d}{2} - 2 \frac{d-4}{d-2a-2}
\label{Beyond:Ia3}\\
&{} = - 4 \frac{a-1}{d-2a-2}
\F{3}{2}{1,\frac{d}{2}-2,\frac{d}{2}-a-1\\\frac{3}{2}d-a-4,\frac{d}{2}-a}{1}
+ A \pi \cot\pi\left(\tfrac{d}{2}-a\right) - 2
\label{Beyond:Ia4}
\end{align}
where
\begin{equation*}
A = \frac{\Gamma(a)\Gamma\left(\frac{3}{2}d-a-4\right)}%
{\Gamma(d-4)\Gamma\left(\frac{d}{2}-1\right)}\,.
\end{equation*}

A curious integral belonging to the current class was considered in~\cite{DHP:90}.
The symmetry allows one to write it in 12 equivalent forms:
\begin{align}
   &\bar{I}(1,1,\lambda,1,\lambda,\lambda)
=   \bar{I}(1,1,\lambda,\lambda,1,\lambda)
=   \bar{I}(1,\lambda,1,1,\lambda,\lambda)
=   \bar{I}(1,\lambda,1,\lambda,\lambda,1)
\nonumber\\
={}&\bar{I}(1,\lambda,\lambda,1,1,\lambda)
=   \bar{I}(1,\lambda,\lambda,1,\lambda,1)
=   \bar{I}(\lambda,1,1,\lambda,1,\lambda)
=   \bar{I}(\lambda,1,1,\lambda,\lambda,1)
\nonumber\\
={}&\bar{I}(\lambda,1,\lambda,1,1,\lambda)
=   \bar{I}(\lambda,1,\lambda,\lambda,1,1)
=   \bar{I}(\lambda,\lambda,1,1,\lambda,1)
=   \bar{I}(\lambda,\lambda,1,\lambda,1,1)
\label{Beyond:Pismak}
\end{align}
where $\lambda=d/2-1$.
It reduces to $I(1,\lambda,\lambda,\lambda)$.
However, it cannot be directly calculated using the above formulas:
one of the arguments should be shifted by $x$,
and the limit $x\to0$ should be taken.
In the paper~\cite{DHP:90},
recurrence relations shifting $d$ by $\pm2$ were derived
(they were used in~\cite{KSV:94}).
This method became popular later.

\section{Mellin--Barnes representation}
\label{S:MB}

The integral $I$ can be written as ($k^2=1$)
\begin{align}
&\raisebox{-4.25mm}{\begin{picture}(22,11)
\put(11,5.5){\makebox(0,0){\includegraphics{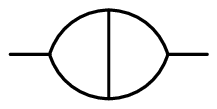}}}
\put(11.5,5.5){\makebox(0,0)[l]{$a_3$}}
\put(6,10){\makebox(0,0){$a_1$}}
\put(16,10){\makebox(0,0){$a_2$}}
\put(16,1){\makebox(0,0){$a_4$}}
\put(6,1){\makebox(0,0){$a_5$}}
\put(11.5,5.5){\makebox(0,0)[l]{$a_3$}}
\end{picture}}
=
\raisebox{-3.75mm}{\begin{picture}(18,10)
\put(9,5){\makebox(0,0){\includegraphics{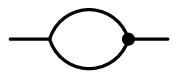}}}
\put(9,9.5){\makebox(0,0){$a_1$}}
\put(9,0.5){\makebox(0,0){$a_5$}}
\end{picture}}
\quad\text{where}\quad
\raisebox{-1.75mm}{\begin{picture}(10,6)
\put(5,3){\makebox(0,0){\includegraphics{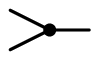}}}
\end{picture}}
=
\raisebox{-4.75mm}{\begin{picture}(16,12)
\put(8,6){\makebox(0,0){\includegraphics{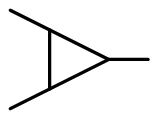}}}
\put(9,9){\makebox(0,0){$a_2$}}
\put(9,3){\makebox(0,0){$a_4$}}
\put(4.5,6){\makebox(0,0)[r]{$a_3$}}
\end{picture}}\,,
\nonumber\\
&I(a_1,a_2,a_3,a_4,a_5) = \frac{1}{\pi^{d/2}}
\int \frac{d^d k_1}{[(k_1-k)^2]^{a_1} (k_1^2)^{a_5}}
V((k_1-k)^2,k_1^2)\,,
\nonumber\\
&V((k_1-k)^2,k_1^2) = \frac{1}{\pi^{d/2}}
\int \frac{d^d k_2}{[(k_2-k)^2]^{a_2} [(k_1-k_2)^2]^{a_3} (k_2^2)^{a_4}}\,.
\label{MB:nest}
\end{align}
Substituting the Mellin--Barnes representation of the one-loop vertex~\cite{BD:91}%
\footnote{For $d=4$ and $a_i=1$ it was obtained in~\cite{U:75}.}
\begin{equation}
V((k_1-k)^2,k_1^2) = \frac{1}{(2\pi i)^2} \int d z_1\,d z_2\,
[(k_1-k)^2]^{z_1} (k_1^2)^{z_2} v(z_1,z_2)
\label{MB:V}
\end{equation}
(where $v(z_1,z_2)$ is a combination of $\Gamma$-functions),
we can easily calculate the loop integral in $k_1$~\cite{BW:03}%
\footnote{For $d=4$ and $a_i=1$ this was also done in~\cite{U:75}.}:
\begin{equation}
I(a_1,a_2,a_3,a_4,a_5) = \frac{1}{(2\pi i)^2} \int d z_1\,d z_2\,
G(a_1-z_1,a_5-z_2) v(z_1,z_2)
\label{MB:I}
\end{equation}
($G(a_1,a_2)$ is the standard massless one-loop self-energy integral).
The result is
\begin{align}
&I(a_1,a_2,a_3,a_4,a_5) = \frac{1}%
{(2\pi i)^2 \Gamma(a_2) \Gamma(a_4) \Gamma(a_3) \Gamma(d-a_2-a_4-a_3)}
\nonumber\\
&{}\times\int d z_1\,d z_2\,
\frac{\Gamma(-z_1) \Gamma(\frac{d}{2}-a_4-a_3-z_1) \Gamma(\frac{d}{2}-a_1+z_1)}%
{\Gamma(a_1-z_1)}
\nonumber\\
&\quad\frac{\Gamma(-z_2) \Gamma(\frac{d}{2}-a_2-a_3-z_2) \Gamma(\frac{d}{2}-a_5+z_2)}%
{\Gamma(a_5-z_2)}
\nonumber\\
&\quad{}\frac{\Gamma(a_1+a_5-\frac{d}{2}-z_1-z_2)
\Gamma(a_3+z_1+z_2) \Gamma(a_2+a_4+a_3-\frac{d}{2}+z_1+z_2)}%
{\Gamma(d-a_1-a_5+z_1+z_2)}\,.
\label{MB:Ires}
\end{align}

This double Mellin--Barnes integral can be expressed via double sums~\cite{BW:03}.
First we close the $z_1$ integration contour to the right.
There are 3 series of poles: $\Gamma(-z_1)$,
$\Gamma(\frac{d}{2}-a_4-a_3-z_1)$, $\Gamma(a_1+a_5-\frac{d}{2}-z_1-z_2)$,
and we obtain 3 sums over residues
\begin{equation*}
I = I_1 + I_2 + I_3\,.
\end{equation*}
Then we close the $z_2$ integration contour contour to the right,
and get double sums:
\begin{equation*}
\begin{split}
I &{}= I_{1\,1} + I_{1\,2} + I_{1\,3}\\
&{} + I_{2\,1} + I_{2\,2} + I_{2\,3}\\
&{} + I_{3\,1} + I_{3\,2} + I_{3\,3} + I_{3\,4} + I_{3\,5}\,.
\end{split}
\end{equation*}

These nested sums belong to the classes which can be expanded in $\varepsilon$
to any order in terms of multiple $\zeta$ values
by the algorithms constructed in~\cite{MUW:02}.
These algorithms were implemented in the packages
\texttt{NestedSums}~\cite{W:02} (in \texttt{C++} with \texttt{GiNaC})
and \texttt{XSUMMER}~\cite{MU:06} (in \texttt{FORM}).
Therefore, expansion of the integrals $I$ to any order in $\varepsilon$
can be written in terms of multiple $\zeta$ values~\cite{BW:03}.

\textbf{Acknowledgements}.
I am grateful to
P.\,A.~Baikov,
D.\,J.~Broadhurst,
K.\,G.~Che\-tyrkin,
A.\,I.~Davydychev,
M.\,Yu.~Kalmykov,
A.\,V.~Kotikov,
V.\,A.~Smirnov
for numerous discussions of various questions related to the present topic;
to Yu.\,M.~Pismak, A.\,P.~Isaev, R.\,N.~Lee, N.\,A.~Kivel
for constructive comments;
to T.~Huber, D.~Ma\^{\i}tre
for their help in using \texttt{HypExp} and \texttt{HPL};
and to D.\,I.~Kazakov and the members of the organizing committee
for organizing the conference and inviting me to present a talk.
This work was supported by the BMBF through Grant No. 05H09VKE.

\appendix
\section{Dispersive calculation of $I(\varepsilon)$}
\label{S:Hathrell}

\begin{figure}[ht]
\begin{center}
\begin{picture}(54,11)
\put(11,5.5){\makebox(0,0){\includegraphics{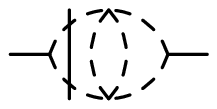}}}
\put(43,5.5){\makebox(0,0){\includegraphics{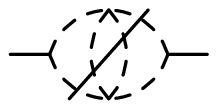}}}
\end{picture}
\end{center}
\caption{Two- and four-particle cuts.}
\label{F:Hathrell}
\end{figure}

The integral
\begin{equation}
\raisebox{-4mm}{\begin{picture}(22,11)
\put(11,5.5){\makebox(0,0){\includegraphics{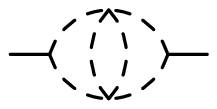}}}
\end{picture}} \sim
\raisebox{-4mm}{\begin{picture}(22,11)
\put(11,5.5){\makebox(0,0){\includegraphics{ibpd.eps}}}
\put(11.5,5){\makebox(0,0)[l]{$\varepsilon$}}
\end{picture}}
\label{Hathrell:def}
\end{equation}
with all indices equal to 1 was considered in~\cite{H:82}.
By dimensionality it is $\sim (k^2)^{-3\varepsilon}$.
At $k^2=-s-i0$ ($s>0$) this factor becomes $s^{-3\varepsilon} e^{3\pi i\varepsilon}$;
its imaginary part is thus $s^{-3\varepsilon} \sin(3\pi\varepsilon)$.
On the other hand, this imaginary part can be calculated
via Cutkosky rules (Fig.~\ref{F:Hathrell}).

The two-particle cuts contribution contains the one-loop triangle
(with one non-integer index) having two legs on-shell;
it is expressible via $\Gamma$-functions.
The $d$-dimensional two-particle phase space also reduces to $\Gamma$-functions.
The integral over the four-particle phase space can be calculated
in terms of ${}_3F_3$ of unit argument, and the result is~\cite{H:82}
\begin{align}
&I(\varepsilon) = (d-4)
\frac{\Gamma\left({\textstyle\frac{d}{2}}-1\right)
\Gamma(d-3) \Gamma\left(5-{\textstyle\frac{3}{2}}d\right)}%
{\Gamma\left(3-\frac{d}{2}\right)}
\label{Hathrell:Ie}\\
&\biggl[ \frac{2 \Gamma\left(\frac{d}{2}-1\right)}{(3d-8) \Gamma(2d-5)}
\F{3}{2}{1,d-2,\frac{3}{2}d-4\\2d-5,\frac{3}{2}d-3}{1}
- \Gamma(3-d) \Gamma\left(2-{\textstyle\frac{d}{2}}\right)
\cos(\pi d)
\biggr]\,.
\nonumber
\end{align}

Using IBP~(\ref{IBP:a3}) we can easily obtain $I(n+\varepsilon)$
for all integer $n$.
However, this method cannot be generalized to $I(n+m\varepsilon)$
with $m\neq1$, unlike~(\ref{Geg:Ia}).

\section{Expanding hypergeometric functions in $\varepsilon$}
\label{S:Hyper}

As we have seen, a large class of integrals $I$ can be expressed
via hypergeometric functions of the argument 1 (or $-1$)
with indices which are integer at $\varepsilon\to0$.
They can be expanded in $\varepsilon$ using the algorithm A from~\cite{MUW:02};
there are also other algorithms~\cite{HM:06,KWY:07}.
It is easy to expand them up to quantities of transcendentality level 8
using the \texttt{Mathematica} package \texttt{HypExp}~\cite{HM:06}
(which uses the package \texttt{HPL}~\cite{M:06}).
As an example, here we expand $\bar{I}(1+\varepsilon)$~(\ref{Sym:barI})
up to $\varepsilon^5$; the result agrees with~(\ref{Sym:e4}), (\ref{Beyond:e5}).
The following text is a \texttt{Mathematica} notebook exported to \LaTeX{}
and slightly edited for readability.

\subsection*{Initialization}

\noindent\(\pmb{<<\text{HypExp$\grave{ }$}}\)
\vspace{2mm}

\noindent\(\text{*-*-*-*-*-* HPL 2.0 *-*-*-*-*-*}\)

\noindent\(\text{Author: Daniel Maitre, University of Zurich}\)

\noindent\(\text{Rules for minimal set loaded for weights: 2, 3, 4, 5, 6, 7, 8.}\)

\noindent\(\text{Rules for minimal set for + - weights loaded for weights: 2, 3, 4, 5, 6, 7, 8.}\)

\noindent\(\text{Table of MZVs loaded up to weight 8}\)

\noindent\(\text{Table of values at I loaded up to weight 7}\)

\noindent\(\text{$\$$HPLFunctions gives a list of the functions of the package.}\)

\noindent\(\text{$\$$HPLOptions gives a list of the options of the package.}\)

\noindent\(\text{More info in hep-ph/0507152, hep-ph/0703052 and at}\)

\noindent\(\text{http://krone.physik.unizh.ch/$\sim $maitreda/HPL/}\)

\noindent\(\text{***********************************}\)

\noindent\(\text{***********  HypExp 2.0  ************}\)

\noindent\(\text{***********************************}\)

\noindent\(\text{Authors:}\)

\noindent\(\text{Tobias Huber:  RWTH Aachen,}\)

\noindent\(\text{Daniel Maitre: SLAC, University of Zurich.}\)



\noindent\(\text{HypExp loaded! It allows the expansion of hypergeometric functions around their parameters.}\)

\noindent\(\text{The new provided commands are:}\)

\noindent\(\text{- HypExp}\)

\noindent\(\text{- HypExpInt}\)

\noindent\(\text{- HypExpU}\)

\noindent\(\text{- HypExpAddToLib}\)

\noindent\(\text{- HypExpIsKnownToOrder}\)

\noindent\(\text{More info in hep-ph/0507094 and at}\)

\noindent\(\text{http://krone.physik.unizh.ch/$\sim $maitreda/HypExp/}\)

\vspace{2mm}
\noindent\(\pmb{\Gamma =\text{Gamma};}\)
\vspace{2mm}

\subsection*{Kotikov}

We start from various expressions for $I(n+\varepsilon)$~(\ref{Geg:Ia})
with $n=0$, 1, 2 as linear expressions in a hypergeometric function $F$, and divide them by
\begin{equation*}
\frac{\Gamma(1+3\varepsilon) \Gamma^2(1-\varepsilon) \Gamma(1-2\varepsilon)}%
{\Gamma(1+\varepsilon) \Gamma(1-4\varepsilon )}\,;
\end{equation*}
then their expansions don't contain the Euler constant $\gamma$,
and the right-hand sides of the IBP relations~(\ref{IBP:a3})
become rational functions.
Then we use these IBP relations to reduce the cases $n=0$, 2 to $n=1$,
and multiply by the appropriate factor to obtain $\bar{I}$~(\ref{Sym:barI}).
The result contains
\begin{equation*}
S_{8a} = \zeta_{5\,3} + \zeta_8\,.
\end{equation*}

\vspace{2mm}
\noindent\(\pmb{i[\text{j$\_$},\text{x$\_$},\text{f$\_$},\text{n$\_$}]\text{:=}\text{Module}[\{a,b,y\},a=D[x,F];b=x\text{/.}F\to 0;b=b/a;}\\
\pmb{a=a*\Gamma [1+\varepsilon ]*\Gamma [1-4*\varepsilon ]/(\Gamma [1+3*\varepsilon ]*\Gamma [1-\varepsilon ]{}^{\wedge}2*\Gamma [1-2*\varepsilon ]);}\\
\pmb{y=\text{Series}[\text{HypExp}[f,\varepsilon ,n],\{\varepsilon ,0,n\}];y=\text{Map}[\text{Expand},\text{FunctionExpand}[a*(y+b)]];}\\
\pmb{y=\text{Switch}[j,0,}\\
\pmb{\text{Map}[\text{Expand},((1-3*\varepsilon )*y-(1-5*\varepsilon )/(3*\varepsilon {}^{\wedge}2*(1-3*\varepsilon )*(1-4*\varepsilon )))/(2*\varepsilon )],}\\
\pmb{2,\text{Map}[\text{Expand},(-(1+2*\varepsilon )*y+(1+5*\varepsilon )/(3*\varepsilon {}^{\wedge}2*(1+\varepsilon )*(1+2*\varepsilon )))/(3*\varepsilon )],}\\
\pmb{\_,y];}\\
\pmb{\text{Map}[\text{Expand},\text{FunctionExpand}[y*(1-2*\varepsilon )*\Gamma [1-\varepsilon ]{}^{\wedge}2*}\\
\pmb{\text{Sqrt}[\Gamma [1-2*\varepsilon ]*\Gamma [1+3*\varepsilon ]/(\Gamma [1+\varepsilon ]*\Gamma [1-4*\varepsilon ])]]]]}\)
\vspace{2mm}

\noindent\(\pmb{\text{FK}[\text{n$\_$}]=\text{HypergeometricPFQ}[\{1,2-2*\varepsilon ,n+2*\varepsilon \},\{n+1+\varepsilon ,n+1+2*\varepsilon \},1];}\)
\vspace{2mm}

\noindent\(\pmb{K[\text{n$\_$}]=2*\Gamma [1-\varepsilon ]*\Gamma [1-n-2*\varepsilon ]*\Gamma [n-1+3*\varepsilon ]*}\\
\pmb{(\pi *\text{Cot}[\pi *(n+3*\varepsilon )]/\Gamma [2-2*\varepsilon ]-}\\
\pmb{\Gamma [1-\varepsilon ]/((n+2*\varepsilon )*\Gamma [2-n-4*\varepsilon ]*\Gamma [n+1+\varepsilon ])*F);}\)
\vspace{2mm}

\noindent\(\pmb{\text{I1}=i[1,K[1],\text{FK}[1],7]}\)

\begin{equation*}
\begin{split}
&6 \zeta (3) + \frac{\pi^4 \varepsilon}{10} + 102 \zeta (5) \varepsilon^2
+ \left(\frac{16 \pi^6}{63} - 24 \zeta(3)^2\right) \varepsilon^3
+ \left(1413 \zeta (7) - \frac{4 \pi^4 \zeta (3)}{5}\right) \varepsilon ^4 + {}\\
&\left(\frac{648 \text{HPLs8a}}{5} - 228 \zeta(3) \zeta(5) + \frac{6017 \pi^8}{15750}\right) \varepsilon^5
+ O\left(\varepsilon^6\right)
\end{split}
\end{equation*}

\noindent\(\pmb{\{i[0,\text{K}[0],\text{FK}[0],8]-\text{I1},i[2,\text{K}[2],\text{FK}[2],6]-\text{I1}\}}\)
\vspace{2mm}

\noindent\(\left\{O\left(\varepsilon^6\right),O\left(\varepsilon^6\right)\right\}\)

\subsection*{Broadhurst, Gracey, Kreimer}

Here we use~(\ref{Beyond:Ia1})--(\ref{Beyond:Ia4}).
In two cases, I was unable to expand up to $\varepsilon^5$: after running for a long time,
\texttt{Mathematica} said ``No more memory available''.
In these cases, I expanded up to $\varepsilon^4$.

\vspace{2mm}
\noindent\(\pmb{b[\text{n$\_$}]\text{:=}-2*\Gamma [1-\varepsilon ]{}^{\wedge}2*\Gamma [n-1+3*\varepsilon ]*}\\
\pmb{\Gamma [1-n-2*\varepsilon ]/(\varepsilon *(1-2*\varepsilon )*\Gamma [n+\varepsilon ]*\Gamma [2-n-4*\varepsilon ])}\)
\vspace{2mm}

\noindent\(\pmb{a[\text{n$\_$}]\text{:=}\Gamma [n+\varepsilon ]*\Gamma [2-n-4*\varepsilon ]/(\Gamma [1-\varepsilon ]*\Gamma [1-2*\varepsilon ])}\)
\vspace{2mm}

\noindent\(\pmb{\text{F1}[\text{n$\_$}]\text{:=}\text{HypergeometricPFQ}[\{1,-\varepsilon ,n-1+3*\varepsilon \},\{n+\varepsilon ,n+3*\varepsilon \},1]}\)
\vspace{2mm}

\noindent\(\pmb{\text{B1}[\text{n$\_$}]\text{:=}b[n]*((1-n-4*\varepsilon )/(1-n-3*\varepsilon )*F-a[n]*\pi *\varepsilon *\text{Cot}[3*\pi *\varepsilon ]-1)}\)
\vspace{2mm}

\noindent\(\pmb{\{i[1,\text{B1}[1]-\text{I1},\text{F1}[1],8],i[0,\text{B1}[0],\text{F1}[0],6]-\text{I1},i[2,\text{B1}[2],\text{F1}[2],7]-\text{I1}\}}\)
\vspace{2mm}

\noindent\(\left\{O\left(\varepsilon^6\right),O\left(\varepsilon^5\right),O\left(\varepsilon^6\right)\right\}\)
\vspace{2mm}

\noindent\(\pmb{\text{F2}[\text{n$\_$}]\text{:=}\text{HypergeometricPFQ}[\{1,1-n-\varepsilon ,1-n-3*\varepsilon \},\{1+\varepsilon ,2-n-3*\varepsilon \},1]}\)
\vspace{2mm}

\noindent\(\pmb{\text{B2}[\text{n$\_$}]\text{:=}b[n]*(-(1-n-4*\varepsilon )/(1-n-3*\varepsilon )*F+a[n]*\pi *\varepsilon *\text{Cot}[\pi *\varepsilon ]-\varepsilon
/(1-n-3*\varepsilon ))}\)
\vspace{2mm}

\noindent\(\pmb{\{i[1,\text{B2}[1],\text{F2}[1],8]-\text{I1},i[0,\text{B2}[0],\text{F2}[0],7]-\text{I1},i[2,\text{B2}[2],\text{F2}[2],7]-\text{I1}\}}\)
\vspace{2mm}

\noindent\(\left\{O\left(\varepsilon^6\right),O\left(\varepsilon^6\right),O\left(\varepsilon^6\right)\right\}\)
\vspace{2mm}

\noindent\(\pmb{\text{F3}[\text{n$\_$}]\text{:=}\text{HypergeometricPFQ}[\{1,n-1+4*\varepsilon ,n-1+2*\varepsilon \},\{1+\varepsilon ,n+2*\varepsilon \},1]}\)
\vspace{2mm}

\noindent\(\pmb{\text{B3}[\text{n$\_$}]\text{:=}b[n]*(-(n-1+\varepsilon )/(n-1+2*\varepsilon )*F+a[n]*\pi *\varepsilon *\text{Cot}[\pi *\varepsilon ]+\varepsilon
/(1-n-2*\varepsilon ))}\)
\vspace{2mm}

\noindent\(\pmb{\{i[1,\text{B3}[1],\text{F3}[1],8]-\text{I1},i[0,\text{B3}[0],\text{F3}[0],7]-\text{I1},i[2,\text{B3}[2],\text{F3}[2],7]-\text{I1}\}}\)
\vspace{2mm}

\noindent\(\left\{O\left(\varepsilon^6\right),O\left(\varepsilon^6\right),O\left(\varepsilon^6\right)\right\}\)
\vspace{2mm}

\noindent\(\pmb{\text{F4}[\text{n$\_$}]\text{:=}\text{HypergeometricPFQ}[\{1,-\varepsilon ,1-n-2*\varepsilon \},\{2-n-4*\varepsilon ,2-n-2*\varepsilon \},1]}\)
\vspace{2mm}

\noindent\(\pmb{\text{B4}[\text{n$\_$}]\text{:=}b[n]*((n-1+\varepsilon )/(n-1+2*\varepsilon )*F+a[n]*\pi *\varepsilon *\text{Cot}[2*\pi *\varepsilon ]-1)}\)
\vspace{2mm}

\noindent\(\pmb{\{i[1,\text{B4}[1],\text{F4}[1],8]-\text{I1},i[0,\text{B4}[0],\text{F4}[0],7]-\text{I1},i[2,\text{B4}[2],\text{F4}[2],6]-\text{I1}\}}\)
\vspace{2mm}

\noindent\(\left\{O\left(\varepsilon^6\right),O\left(\varepsilon^6\right),O\left(\varepsilon^5\right)\right\}\)

\subsection*{Hathrell}

For this particular problem we can also use~(\ref{Hathrell:Ie}).

\vspace{2mm}
\noindent\(\pmb{\text{FH}=\text{HypergeometricPFQ}[\{1,2-2*\varepsilon ,2-3*\varepsilon \},\{3-4*\varepsilon ,3-3*\varepsilon \},1];}\)
\vspace{2mm}

\noindent\(\pmb{H=-2*\varepsilon *\Gamma [1-\varepsilon ]*\Gamma [-1+3*\varepsilon ]*\Gamma [1-2*\varepsilon ]/\Gamma [1+\varepsilon ]*}\\
\pmb{(\Gamma [1-\varepsilon ]/((2-3*\varepsilon )*\Gamma [3-4*\varepsilon ])*F-\Gamma [-1+2*\varepsilon ]*\Gamma [\varepsilon ]*\text{Cos}[2*\pi *\varepsilon ]);}\)
\vspace{2mm}

\noindent\(\pmb{i[0,H,\text{FH},6]-\text{I1}}\)
\vspace{2mm}

\noindent\(O\left(\varepsilon ^6\right)\)

\subsection*{Kazakov}

Expanding hypergeometric functions of $-1$ is a little tricky.
First, we should switch off automatic conversion of harmonic polylogarithms
to usual polylogarithms
(otherwise values of polylogarithms on their cuts will be generated).
These expansions involve, in addition to multiple $\zeta$ values,
alternating (Euler--Zagier) sums.
After reducing them to a minimal set,
many new constants appear (including $\log2$ and polylogarithms of $\frac{1}{2}$).
These new constants cancel in the square bracket in~(\ref{Uni:F32})
(it is calculated by the function \textbf{br}).

\vspace{2mm}
\noindent\(\pmb{\text{$\$$HPLAutoConvertToKnownFunctions}=\text{False};}\)
\vspace{2mm}

\noindent\(\pmb{\text{FK1}[\text{n$\_$}]\text{:=}\text{HypergeometricPFQ}[\{1,2-2*\varepsilon ,n+2*\varepsilon \},\{1+\varepsilon ,n+1+2*\varepsilon \},-1]}\)
\vspace{2mm}

\noindent\(\pmb{\text{FK2}[\text{n$\_$}]\text{:=}\text{HypergeometricPFQ}[\{1,2-2*\varepsilon ,2-n-3
*\varepsilon \},\{1+\varepsilon ,3-n-3*\varepsilon \},-1]}\)
\vspace{2mm}

\noindent\(\pmb{\text{br}[\text{j$\_$},\text{n$\_$}]\text{:=}\text{Module}[\{\text{f1},\text{f2},
\text{n1},\text{n2},y\},\text{n1}=n-\text{If}[j\text{===}2,1,0];}\\
\pmb{\text{n2}=n-\text{If}[j\text{===}0,1,0];}\\
\pmb{\text{f1}=\text{Map}[\text{Expand},\text{Series}[\text{HypExp}[\text{FK1}[j],\varepsilon ,\text
{n1}],\{\varepsilon ,0,\text{n1}\}]];}\\
\pmb{\text{f2}=\text{Map}[\text{Expand},\text{Series}[\text{HypExp}[\text{FK2}[j],\varepsilon ,\text
{n2}],\{\varepsilon ,0,\text{n2}\}]];}\\
\pmb{\text{Map}[\text{Expand},\text{f1}/(j+2*\varepsilon )+\text{f2}/(2-j-3*\varepsilon )]]}\)
\vspace{2mm}

\noindent\(\pmb{\text{ik}[\text{j$\_$},\text{n$\_$}]\text{:=}\text{Module}[\{y\},y=\text{br}[j,n];}\\
\pmb{y=\text{Map}[\text{Expand},}\\
\pmb{\text{FunctionExpand}[y-\Gamma [1+\varepsilon ]*\Gamma [j+\varepsilon ]*\Gamma [2-j-4*\varepsilon ]/\Gamma [2-2*\varepsilon ]*\text{Cos}[\pi *\varepsilon ]]];}\\
\pmb{y=\text{Map}[\text{Expand},2*\text{Pochhammer}[1+3*\varepsilon ,j-2]*}\\
\pmb{\text{Pochhammer}[1-2*\varepsilon ,-j]/(\text{Pochhammer}[\varepsilon ,j]*\text{Pochhammer}[1-4*\varepsilon ,1-j])*y];}\\
\pmb{y=\text{Switch}[j,0,}\\
\pmb{\text{Map}[\text{Expand},((1-3*\varepsilon )*y-(1-5*\varepsilon )/(3*\varepsilon {}^{\wedge}2*(1-3*\varepsilon )*(1-4*\varepsilon )))/(2*\varepsilon )],}\\
\pmb{2,\text{Map}[\text{Expand},(-(1+2*\varepsilon )*y+(1+5*\varepsilon )/(3*\varepsilon {}^{\wedge}2*(1+\varepsilon )*(1+2*\varepsilon )))/(3*\varepsilon )],}\\
\pmb{\_,y];}\\
\pmb{\text{Map}[\text{Expand},\text{FunctionExpand}[y*(1-2*\varepsilon )*\Gamma [1-\varepsilon ]{}^{\wedge}2*}\\
\pmb{\text{Sqrt}[\Gamma [1-2*\varepsilon ]*\Gamma [1+3*\varepsilon ]/(\Gamma [1+\varepsilon ]*\Gamma [1-4*\varepsilon ])]]]]}\)
\vspace{2mm}

\noindent\(\pmb{\{\text{ik}[1,8]-\text{I1},\text{ik}[0,8]-\text{I1},\text{ik}[2,8]-\text{I1}\}}\)
\vspace{2mm}

\noindent\(\left\{O\left(\varepsilon^6\right),O\left(\varepsilon^6\right),O\left(\varepsilon^6\right)\right\}\)

\end{document}